\documentclass[10pt]{article}
\pdfoutput=1
\usepackage[utf8x]{inputenc}
\usepackage{amssymb,amsmath,mathrsfs,enumerate}
\usepackage{rotate,multicol}
\usepackage{graphicx,wrapfig}
\usepackage{tikz,graphics,color,fullpage,float,epsf,subcaption}
\usepackage{pdfpages}
\usepackage{mathtools}
\usepackage[textfont=it,labelfont=bf]{caption}
\usepackage[colorlinks=true,
  linkcolor=red,
  urlcolor=blue,
  citecolor=blue]{hyperref}
\usepackage{color}
\usepackage{soul}
\usepackage{cite}
\usepackage{verbatim}
\usepackage[symbol]{footmisc}
\long\def\rpl#1!!#2!!{\textcolor{red}{#1} \textcolor{blue}{#2}}

\DeclareMathOperator{\Tr}{Tr}

\def \order(#1){{\cal O} \left(#1 \right)}

\newmuskip\pFqskip
\mathchardef\pFcomma=\mathcode`, 

\evensidemargin=\oddsidemargin
\def\Eqn#1{Eq.\ (\ref{#1})}
\def\Eqs#1#2{Eqs.\ (\ref{#1}) and (\ref{#2})}
\allowdisplaybreaks
\begin{document}
\begin{flushright}
\mbox{}
TIFR/TH/23-18\\
\end{flushright}


\vskip 30pt

\begin{center}
  {\Large \bf New physics implications of VBF searches
    
    exemplified through the Georgi-Machacek model} \\
  \vspace*{1cm} {\sf Manimala
    Chakraborti$^{a,}$\footnote[1]{mani.chakraborti@gmail.com},~Dipankar
    Das$^{b,}$\footnote[2]{d.das@iiti.ac.in},~Nivedita
    Ghosh$^{c,}$\footnote[4]{niveditag@iisc.ac.in},~Samadrita 
    Mukherjee$^{d,}$\footnote[3]{samadrita.mukherjee@tifr.res.in},~Ipsita
    Saha$^{e,}$\footnote[5]{ipsita@iitm.ac.in}} \\
  \vspace{10pt} {\small \em 
       $^a$ School of Physics and Astronomy, University of Southampton, Southampton, SO17 1BJ, United Kingdom \\
   $^b$Department of Physics, Indian Institute of Technology(Indore), Khandwa Road, Simrol, 453 552 Indore, India \\
    $^c$Centre for High Energy Physics, Indian Institute of Science,  
    Bengaluru 560012, India\\
    $^d$Department of Theoretical Physics, Tata Institute of Fundamental Research, Mumbai 400005, India \\
    $^e$Department of Physics, Indian Institute of Technology Madras, Chennai 600036, India}
  
  \normalsize
\end{center}

\renewcommand*{\thefootnote}{\arabic{footnote}}
\setcounter{footnote}{0} 
\begin{abstract}

  LHC searches for nonstandard scalars in vector boson fusion (VBF)
  production processes can be particularly efficient in probing scalars belonging to
  triplet or higher multiplet representations of the Standard Model $SU(2)_L$ gauge group.
  They can be especially relevant for models where the additional scalars do not have any tree level couplings
  to the Standard Model fermions, rendering VBF as their primary production mode at the
  LHC. In this work we employ the latest LHC data from VBF resonance searches to
  constrain the properties of nonstandard scalars, taking the
  Georgi-Machacek model as a prototypical example. We take into account the theoretical
  constraints on the potential from unitarity and boundedness-from-below
  as well as indirect constraints coming from the signal strength
  measurements of the 125 GeV Higgs boson at the LHC.
  To facilitate the phenomenological analysis we advocate a convenient reparametrization of the
  trilinear couplings in the scalar potential. We derive simple correlations among the model parameters
  corresponding to the decoupling limit of the model. We explicitly demonstrate how a combination of
  theoretical and phenomenological constraints can push the GM model towards the decoupling limit.
  Our analysis suggests that the VBF searches can provide
  key insights into the composition of the electroweak vacuum expectation value.

\end{abstract}

\bigskip
\newpage
\section{Introduction}
\label{s:intro}
The remarkable success of the Standard Model (SM), culminating in
the discovery of the 125 GeV Higgs scalar at the LHC, has fuelled further
investigations into the understanding of the precise nature of electroweak
symmetry breaking (EWSB). In the $SU(2)_L \times U(1)_Y$ gauge theory of the SM,
the EWSB is driven by an $SU(2)_L$ scalar doublet in such a way that 
the electric charge $Q$, with $Q = T_{3L} + \frac{Y}{2}$, remains conserved~\cite{Gunion:1989we,Djouadi:2005gi}.
A question that naturally arises in this context
is whether there exist additional scalar multiplets of $SU(2)_L$ beyond the SM (BSM)
that contribute to the mechanism of EWSB. If such a scenario is indeed realized in nature,
the nonstandard scalars originating from an extended scalar sector
are expected to possess trilinear couplings with a pair of massive SM gauge
bosons (of the form $S V_1^\mu V_{2 \mu}$) with strengths
proportional to the vacuum expectation values (VEVs) of the extra multiplets.
Thus, a straightforward way to look for such nonstandard scalars
at the LHC is via their production in vector boson fusion (VBF) production processes, provided
the BSM scenario in question can accommodate sizable trilinear couplings of the scalars to a pair
of weak gauge bosons. In other words, the LHC searches for nonstandard
scalar resonances in VBF production processes can
potentially serve as a powerful tool to pin down any BSM contribution to the
process of EWSB.

Nonstandard contribution to the electroweak VEV can arise only from the presence of scalar
multiplets transforming nontrivially under the $SU(2)_L$ part of the SM,
since $SU(2)_L$ singlet scalars, even if charged under the hypercharge gauge group $U(1)_Y$,
do not participate in EWSB. In the popular multi-Higgs doublet extensions
of the SM Higgs sector, the tree level trilinear couplings of the nonstandard Higgs bosons to a pair
of weak gauge bosons vanish in the well-known `alignment limit' of the model~\cite{Grifols:1980uq,Das:2019yad}.
In this limit of multi-Higgs doublet models, the Higgs doublet fields
can be rotated to the so-called Higgs basis where
the electroweak VEV can be effectively allocated entirely to only one of the Higgs doublets,
while the VEVs of all other doublets remain zero~\cite{Branco:2011iw,Bhattacharyya:2015nca}.
The chosen Higgs doublet with non-zero VEV can then give rise to the SM-like Higgs boson observed at the LHC~\cite{ATLAS:2012yve,CMS:2012qbp}.
Thus, the nonstandard scalars in the alignment limit do not possess any trilinear
coupling to the weak gauge boson pairs.
With the current experimental data pointing strongly towards the validity of the alignment limit~\cite{ATLAS:2021vrm}, the possibility of a significantly large trilinear coupling of the nonstandard
scalars to a pair of weak vector bosons
seems to be thin on the ground. One must note here that the introduction of additional
scalar singlets and doublets to the SM field content preserves the custodial $SU(2)$
symmetry and is therefore safe from the stringent constraints stemming from the precise
measurement of electroweak $\rho$-parameter.

Moving beyond the doublets, the next step will be to introduce an additional triplet scalar field into the SM
scalar sector,
giving rise to the so-called Higgs Triplet Model (HTM). Unlike the BSM scenarios with extra singlets and doublets,
those involving triplets or higher multiplets of $SU(2)_L$, in general, have the tendency to modify
the tree level value of the electroweak $\rho$-parameter.
For the HTM, the triplet VEV can significantly alter the tree level value of electroweak $\rho$-parameter, making
the model vulnerable to the constraints coming from precision electroweak measurements. The experimental
determination of the electroweak $\rho$-parameter, $\rho \approx 1$,
severely restricts the VEV of the neutral component of the triplet to values less than a few
GeV~\cite{Konetschny:1977bn,Cheng:1980qt,Kanemura:2012rs}.
This tones down any potential enhancement in the trilinear couplings of the form $S V_1^\mu V_{2 \mu}$
which are directly proportional to the triplet VEVs~\cite{Chang:2012gn}, thereby reducing their observability
in the high energy collider experiments.
Thus, in order to accommodate a substantially large triplet VEV in a
BSM scenario  the vacuum state of the model
must be designed in such a way that it preserves the custodial symmetry.
A prototypical theoretical framework
within this class of constructions is provided by the Georgi-Machacek (GM)
model~\cite{Georgi:1985nv,Chanowitz:1985ug,Gunion:1989ci}.
In the GM model, the SM scalar sector
is extended by introducing two additional triplet scalar fields (one complex and one real)
which acquire equal VEVs. This leads to an internal cancellation
of the custodial symmetry breaking effects, securing $\rho =1$ at tree level.
As a consequence of the unbroken custodial symmetry,
the common triplet VEVs can now be sizable, allowing for significantly large trilinear couplings
of the nonstandard Higgs bosons to the massive gauge boson pairs. Thus, the GM model can serve as the
ideal candidate to illustrate the potential of VBF searches at the LHC 
in determining the basic ingredients of the electroweak VEV.

Several LHC searches have been designed specifically to
look for VBF production of nonstandard scalars in
diboson final states~\cite{ATLAS:2020fry,ATLAS:2020tlo,ATLAS:2021jol,ATLAS:2022jho,ATLAS:2022qlc}.
In this work, we use the latest results from the full LHC Run-II data set
to constrain the properties of additional triplet scalars, within the example framework of GM model.
The GM model consists of several nonstandard scalar bosons
which can give rise to very distinctive phenomenologies at the colliders.
The physical scalar spectrum of the model can be categorized into a number of
custodial multiplets,
namely, one fiveplet $(H_5^{++},H_5^+,H_5^0, H_5^-, H_5^{--})$,
one triplet $(H_3^+,H_0, H_3^-)$ and two singlets $h$ and $H$.
One important feature of this model is that the members of the
custodial fiveplet do not couple to the SM fermions at tree level.
Thus, the dominant direct bounds on the common mass of the custodial fiveplet
can be obtained from the LHC searches looking for nonstandard
bosons in VBF production processes, making these searches
tailor-made for this model. 
In addition to collider data, we also consider the theoretical constraints from perturbative unitarity and 
boundedness-from-below (BFB) conditions on the potential.
A suitable reparametrization of the trilinear couplings of the scalar potential is prescribed,
making way for efficient phenomenological analysis.
We show that the decoupling limit of the GM model can be expressed in terms
of simple correlations among the physical masses, mixing angles and the triplet VEV.
We also take into consideration the bounds coming from the
measurement of the properties of 125 GeV Higgs boson, which we identify as the lightest
CP-even custodial singlet boson $h$ present in the GM model.
In our analysis, we demonstrate that the VBF searches can provide
complementary constraints to the theoretical bounds on the model parameter space.
We also give estimates for the potential of the future colliders such as the 
high luminosity- (HL-) LHC, Future Circular colliders (FCC) etc.
to constrain the remaining parameter space further.
We explicitly show that the projected limits from the HL-LHC VBF searches
for new scalar resonances in combination with the theoretical
constraints will practically push the GM model towards the decoupling limit,
imposing stringent constraints on the triplet contribution to the electroweak VEV.

Analyses trying to constrain the GM model parameter space from the LHC Run-II
data have been performed earlier in the literature~\cite{Chiang:2013rua,Englert:2013zpa,Englert:2013wga,Chiang:2014bia,
  Chiang:2018cgb,Das:2018vkv,Ghosh:2019qie,Ismail:2020zoz,Ismail:2020kqz,Chen:2022zsh,Wang:2022okq,Ghosh:2022wbe,
Bairi:2022adc,deLima:2022yvn,Ghosh:2022bvz,Ahriche:2022aoj}.
Our analysis supersedes them by including the full Run-II data set for the diboson
resonance searches, some of which were not available during the time of the previous analyses.
Apart from the direct collider searches, we employ the latest measurement of the trilinear
Higgs self-coupling to provide complementary
constraints on the GM model parameter space.
Another important feature of our analysis is the simple parametrization
of the input variables in terms of physical quantities resulting in
a more direct interpretation of the phenomenological results.

The plan of our paper is as follows. In Sec.~\ref{s:model}, we briefly
review the GM model. In Sec.~\ref{s:uni}, we discuss the theoretical constraints
on the model from perturbative unitarity and BFB requirements of the potential.
Here we also formulate correlations among the physical parameters corresponding
to the decoupling limit of the model.
Constraints from the 125 GeV Higgs signal strength measurements are described in Sec.~\ref{s:mu}.
In Sec.~\ref{s:direct} we describe the impact of the direct search constraints
from the LHC on the parameter space of the GM model. The prospect for future colliders like the 
HL-LHC to further constrain this scenario is discussed in Sec.~\ref{s:future}. Finally, we summarize
our findings in Sec.~\ref{s:summary}.

\section{A brief recap of the GM Model}
\label{s:model}
The GM model extends the scalar sector of the SM, consisting of the $Y=1$ complex doublet
$\phi \equiv (\phi^+ \quad \phi^0)^\intercal$, by adding two $SU(2)_L$ triplet scalar fields,
one real $\xi \equiv (\xi^+ \quad \xi^0 \quad \xi^-)^\intercal$
and one complex $\chi \equiv (\chi^{++} \quad \chi^+ \quad \chi^0)^\intercal$,
with hypercharges $Y=0$  and $Y=2$ respectively
~\cite{Georgi:1985nv,Chanowitz:1985ug,Gunion:1989ci}. The scalar sector of this model
is conventionally expressed in terms of a bi-doublet $\Phi$ and a bi-triplet $X$, defined as,
\begin{eqnarray}
\label{eq:fields}
\Phi = \left(\begin{array}{cc}
\phi^{0*} & \phi^+ \\
-\phi^- & \phi^0 \\
\end{array}\right) \,, \qquad
X= \left(\begin{array}{ccc}
\chi^{0*} & \xi^+ & \chi^{++} \\
-\chi^- & \xi^0 & \chi^+ \\
\chi^{--} & -\xi^- & \chi^0 \\
\end{array}\right) \,.
\end{eqnarray}
The scalar potential for this model can be written as~\cite{Hartling:2014zca,Das:2018vkv}
\begin{eqnarray}
V(\Phi,X) &=& \frac{\mu_{\phi}^2}{2} \Tr(\Phi^\dagger \Phi) + \frac{\mu_{X}^2}{2} \Tr(X^\dagger X) + \lambda_1[\Tr(\Phi^\dagger \Phi)]^2 +
\lambda_2 \Tr(\Phi^\dagger \Phi)\Tr(X ^\dagger X) \nonumber \\ 
&& + \lambda_3 \Tr(X^\dagger X X^\dagger X) + \lambda_4 [\Tr(X^\dagger X)]^2 - \lambda_5 \Tr(\Phi^\dagger \tau_a \Phi \tau_b)\Tr(X^\dagger t_a X t_b) \nonumber \\
&& - M_1\Tr(\Phi^\dagger \tau_a \Phi \tau_b)\left( U X U^\dagger \right)_{ab} - M_2\Tr(X^\dagger t_a X t_b)\left( U X U^\dagger \right)_{ab} \,,
\label{eq:potential}
\end{eqnarray}
with $\tau_a \equiv \sigma_a/2$, ($a=1,2,3$) where $\sigma_a$'s  are  
the Pauli matrices and $t_a$'s are the generators of the triplet representation of
$SU(2)_L$ and are given by,
\begin{eqnarray}
t_1 = \frac{1}{\sqrt{2}}\left(\begin{array}{ccc}
0 & 1 & 0 \\
1 & 0 & 1 \\
0 & 1 & 0 \\
\end{array}\right)\,, \qquad
t_2 = \frac{1}{\sqrt{2}}\left(\begin{array}{ccc}
0 & -i & 0 \\
i & 0 & -i \\
0 & i & 0 \\
\end{array}\right)\,, \qquad
t_3 =\left(\begin{array}{ccc}
1 & 0 & 0 \\
0 & 0 & 0 \\
0 & 0 & -1 \\
\end{array}\right)\,.
\label{su2_tripgen}
\end{eqnarray}
The matrix $U$ appearing in the trilinear terms of \Eqn{eq:potential} is given by,
\begin{eqnarray}
U = \frac{1}{\sqrt{2}} \left(\begin{array}{ccc}
-1 & 0 & 1 \\
-i & 0 & -i \\
0 & \sqrt{2} & 0 \\
\end{array}\right) \,.
\label{matU}
\end{eqnarray}
After the EWSB, the neutral components of the
bi-doublet and the bi-triplet are expanded around their VEVs as,
\begin{equation}
\phi^0 = \frac{1}{\sqrt{2}} (v_d + h_d + i \eta_d)\;, \quad \xi^0 =  (v_t + h_\xi ) \;, \quad
\chi^0 =  \left(v_t + \frac{h_\chi + i \eta_\chi}{\sqrt{2}}\right).
\label{eq:vevs}
\end{equation}
The requirement of equal VEVs to the real and the complex triplets
ensures that custodial symmetry in the scalar potential remains intact.
From the expressions of $W$ and $Z$ boson masses, the electroweak
VEV can be identified as
\begin{eqnarray}
\sqrt{v_d^2 + 8 v_t^2} = v = 246~{\rm GeV} \,.
\label{eq:custodial}
 \end{eqnarray}
Thus, there will be two independent minimization conditions for the scalar potential
corresponding to the two VEVs of the bi-doublet and the bi-triplet ($v_d$ and $v_t$).
These can be used to extract the bilinear coefficients of the potential $\mu_{\phi}^2$ and $\mu_X^2$ in terms
of $v_d$ and $v_t$ as follows,
\begin{subequations}
	\begin{eqnarray}
	\mu_{\phi}^2 &=&  - 4 \lambda_1 v_d^2 - 3\left(2\lambda_2 -\lambda_5\right) v_t^2 + \frac{3}{2} M_1 v_t \,, \\
	\mu_{X}^2 &=&   -\left(2\lambda_2 -\lambda_5\right) v_d^2 - 4 \left(
	\lambda_3+ 3\lambda_4 \right)v_t^2 + \frac{M_1 v_d^2}{4v_t} + 6 M_2 v_t \,.
	\end{eqnarray}
	\label{eq:bilinears}
\end{subequations}
Now, the bilinear terms in the scalar potential can be diagonalized
to obtain the physical Higgs scalars of the model which can be classified according to
their transformation properties under the custodial SU(2) as a
quintuplet $(H_5^{++},H_5^+,H_5^0, H_5^-, H_5^{--})$ with common mass
$m_5$, a triplet $(H_3^+,H_0, H_3^-)$ of common mass $m_3$ and two custodial singlets, $h$ and $H$ with masses 
$m_h$ and $m_H$ respectively.
In this article, we refrain ourselves from giving a detailed description
of the diagonalization procedure and we refer the reader to
Refs.~\cite{Chiang:2012cn,Hartling:2014zca}. The mass eigenstates
for the charged and neutral scalars are defined below:\footnote{Note that our convention of $\alpha$
differs from that of Ref.~\cite{Hartling:2014zca} by a negative sign.}
\begin{subequations}
\begin{eqnarray}
H_5^{\pm\pm}   &=& \chi^{\pm\pm}, \\
H_5^\pm &=& \frac{1}{\sqrt{2}} \left(\chi^\pm - \xi^\pm \right) \,,  \\
H_5^0 &=& \sqrt{\frac{2}{3}} h_\xi - \sqrt{\frac{1}{3}} h_\chi, \nonumber \\
H_3^\pm &=& -\sin\beta~ \phi^\pm +  \frac{\cos \beta}{\sqrt{2}} \left(\chi^\pm + \xi^\pm \right) \,, \\
H_3^0   &=& -\sin\beta~\eta_d + \cos\beta~\eta_\chi, \\
\label{e:h}
h &=& \cos\alpha~ h_d + \sin\alpha~ H_5^{0\prime} \,, \\
H &=& -\sin\alpha ~h_d + \cos\alpha~ H_5^{0\prime} \,,
\end{eqnarray}
\label{eq:ch}
\end{subequations}
where 
\begin{eqnarray}
H_5^{0\prime} &=& \sqrt{\frac{1}{3}}h_\xi + \sqrt{\frac{2}{3}}h_\chi \,.
\label{eq:h5p0}
\end{eqnarray}
The angle $\alpha$ represents the mixing angle in the neutral Higgs sector
while $\tan\beta$ is defined as
\begin{eqnarray}
\label{e:tanb}
\tan\beta = \frac{2\sqrt{2} v_t}{v_d} \,.
\end{eqnarray}
It can be observed from \Eqn{eq:ch} that the members of the custodial fiveplet
are composed entirely of $SU(2)_L$ scalar triplets without any admixture from
the doublets. Considering the fact that the SM fermions can couple
only to the doublet component, the members of the custodial fiveplet will not have
any tree level coupling to the SM fermions. Thus, the dominant production mode available
for these particles at the LHC is via the VBF process, making the VBF searches
essential to probe the properties of such particles.

Before closing this section we note from \Eqn{eq:potential} that there are nine parameters in the
GM scalar potential with two bilinears ($\mu_{\phi}^2$ and $\mu_X^2$),
five quartic couplings ($\lambda_i$, $i=1,\dots, 5$) and two trilinear
couplings ($M_1$ and $M_2$). Among these, the bilinears can be replaced by the VEVs, $v_d$ and $v_t$
using \Eqn{eq:bilinears}. The five quartic couplings can also be
exchanged for the four physical scalar masses, 
$m_5$, $m_3$, $m_H$ and $m_h$ and the mixing angle, $\alpha$.
Below, we present the relation between the $\lambda_i$-s with the physical
masses and mixings~\cite{Das:2018vkv} :
\begin{subequations}
\label{e:lambdas}
	\begin{eqnarray}
	\lambda_1 &=& \frac{1}{8 v^2 \cos^2 \beta}\left(m_h^2 \cos^2 \alpha +m_H^2 \sin^2 \alpha\right) \,, \label{lambda1} \\
		\lambda_2 &=&\frac{1}{12 v^2 \cos \beta \sin \beta}\left( \sqrt{6}\left(m_h^2  - m_H^2\right) \sin 2 \alpha + 12 m_3^2 \sin\beta \cos\beta - 3 \sqrt{2}v \cos\beta M_1 \right) \,, \label{lambda2}\\	
\lambda_3 &=& \frac{1}{v^2 \sin^2 \beta}\left( m_5^2 - 3 m_3^2 \cos^2\beta + \sqrt{2}v \cos\beta \cot \beta M_1 - 3 \sqrt{2} v \sin\beta M_2  \right)  \,, \label{lambda3} \\
\lambda_4 &=& \frac{1}{6 v^2 \sin^2 \beta}\Big( 2 m_H^2 \cos^2 \alpha + 2 m_h^2 \sin^2\alpha - 2 m_5^2 
+6 \cos^2\beta m_3^2  - 3\sqrt{2} v \cos\beta \cot \beta M_1 \nonumber \\
&&    + 9 \sqrt{2} v \sin \beta  M_2 \Big) \,, \label{lambda4} \\
	\lambda_5 &=&  \frac{2 m_3^2}{v^2} -\frac{ \sqrt{2} M_1}{ v \sin \beta} \,. \label{lambda5}
		\end{eqnarray}
		\label{eq:masstolam}
\end{subequations}

\noindent
We wish to reiterate that in our analysis we consider $h$ to be the lightest $CP$-even scalar
corresponding to the Higgs boson discovered at the LHC with mass $m_h \approx 125$~GeV.
In this study we will focus on the VBF
production of the nonstandard GM scalars at the LHC.

\section{Theoretical constraints and the decoupling limit}
\label{s:uni}
To motivate the benchmark choices for our phenomenological analysis later, it is
important to discuss the implications of the theoretical constraints from
tree-unitarity and BFB\cite{Aoki:2007ah,Hartling:2014zca}. We will
present our observations in terms of the physical masses and mixings and focus
on formulating a systematic method to approach the SM-like limit a.k.a. the decoupling
limit\cite{Hartling:2014zca}. It is quite intuitive that the decoupling limit will be achieved
when $v_t\ll v$ and all the nonstandard scalars are much heavier than the electroweak
scale. Since this has been already discussed in Ref.~\cite{Hartling:2014zca}, we will be brief
and report only the important relations relevant to our present study. The distinct
upshot of our analysis is that the relations we obtain involve only the physical parameters and therefore are quite straightforward to implement into the numerical
codes, giving us a greater control over the parameters required for the phenomenological studies.

To begin with, we suggestively reparametrize the trilinear coupling parameters
$M_1$ and $M_2$ as follows:
\begin{subequations}
\label{e:Lam}
\begin{eqnarray}
	\label{e:M1}
	\Lambda_1^2 &=& \frac{M_1 v}{\sqrt{2}\sin\beta} \equiv \frac{M_1 v^2}{4v_t} \,, \\
	\Lambda_2^2 &=& 3\sqrt{2}\, v\, M_2\sin\beta \equiv 12\, v_t  M_2 \,.
\end{eqnarray}
\end{subequations}
With these reparametrizations let us now investigate the unitarity conditions.
Theoretical constraints from perturbative unitarity put upper bounds
on the eigenvalues of the $2\rightarrow 2$ scalar scattering amplitude matrix.
The eigenvalues can be expressed in
terms of certain independent combinations of the scalar quartic couplings,
given as~\cite{Aoki:2007ah,Hartling:2014zca},
\begin{subequations}
\label{eq:eigenuni}
\begin{eqnarray}
x_1^{\pm} &=& 12\lambda_1 + 14\lambda_3 + 22\lambda_4 \pm
\sqrt{\left(12\lambda_1 - 14\lambda_3 - 22\lambda_4\right)^2 +144\lambda_2^2} \, , \\
x_2^{\pm} &=& 4\lambda_1 - 2\lambda_3 + 4\lambda_4 \pm
\sqrt{\left(4\lambda_1 + 2\lambda_3 - 4\lambda_4\right)^2 + 4\lambda_5^2} \, , \\
y_1 &=& 16\lambda_3 + 8\lambda_4 \, , \\
y_2 &=& 4\lambda_3 + 8\lambda_4 \, , \\
y_3 &=& 4\lambda_2 - \lambda_5 \, , \\
y_4 &=& 4\lambda_2 + 2\lambda_5 \, , \\
y_5 &=& 4\lambda_2 - 4\lambda_5 \,.
\end{eqnarray}
\end{subequations}
The theoretical constraints from perturbative unitarity
requires that each of these eigenvalues must obey the condition $|x_i^\pm|, |y_i| \le 8\pi$.

\begin{figure}
\centering
\includegraphics[width=0.48\textwidth,height=0.25\textheight]{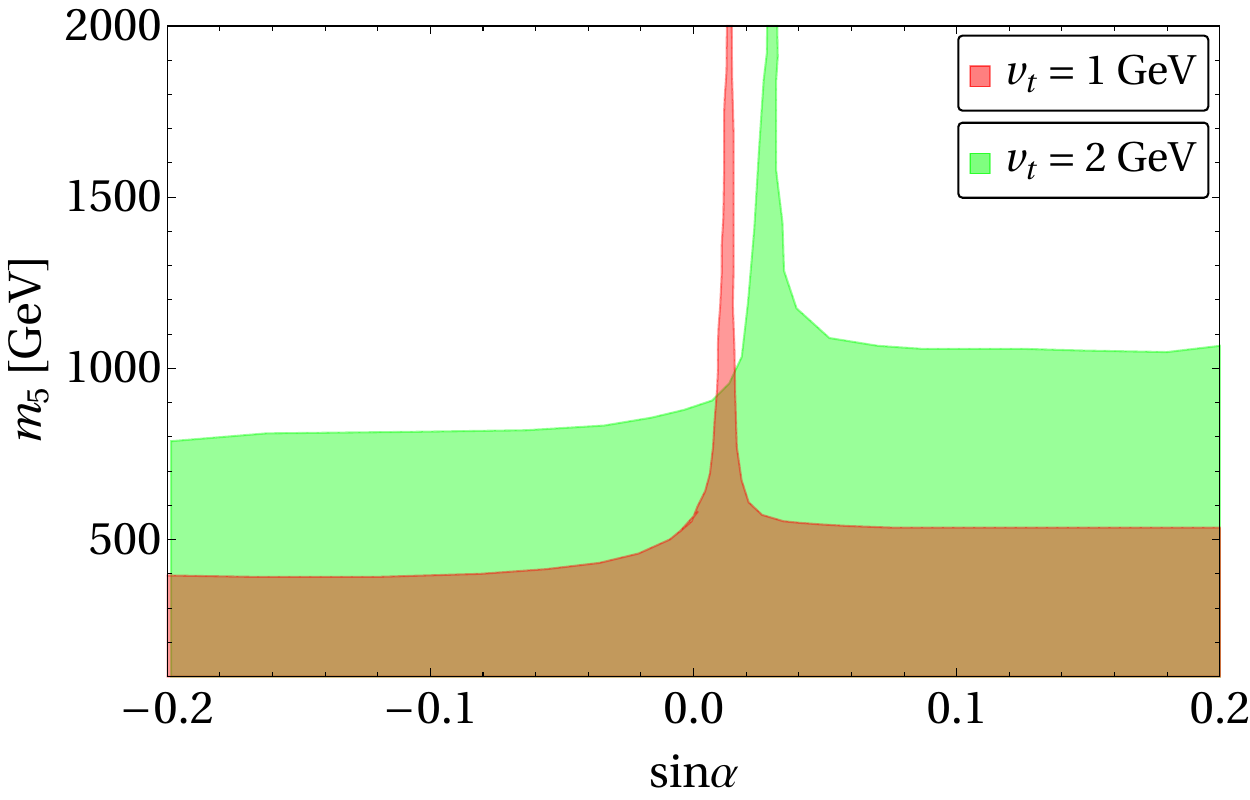} \quad
\includegraphics[width=0.48\textwidth,height=0.25\textheight]{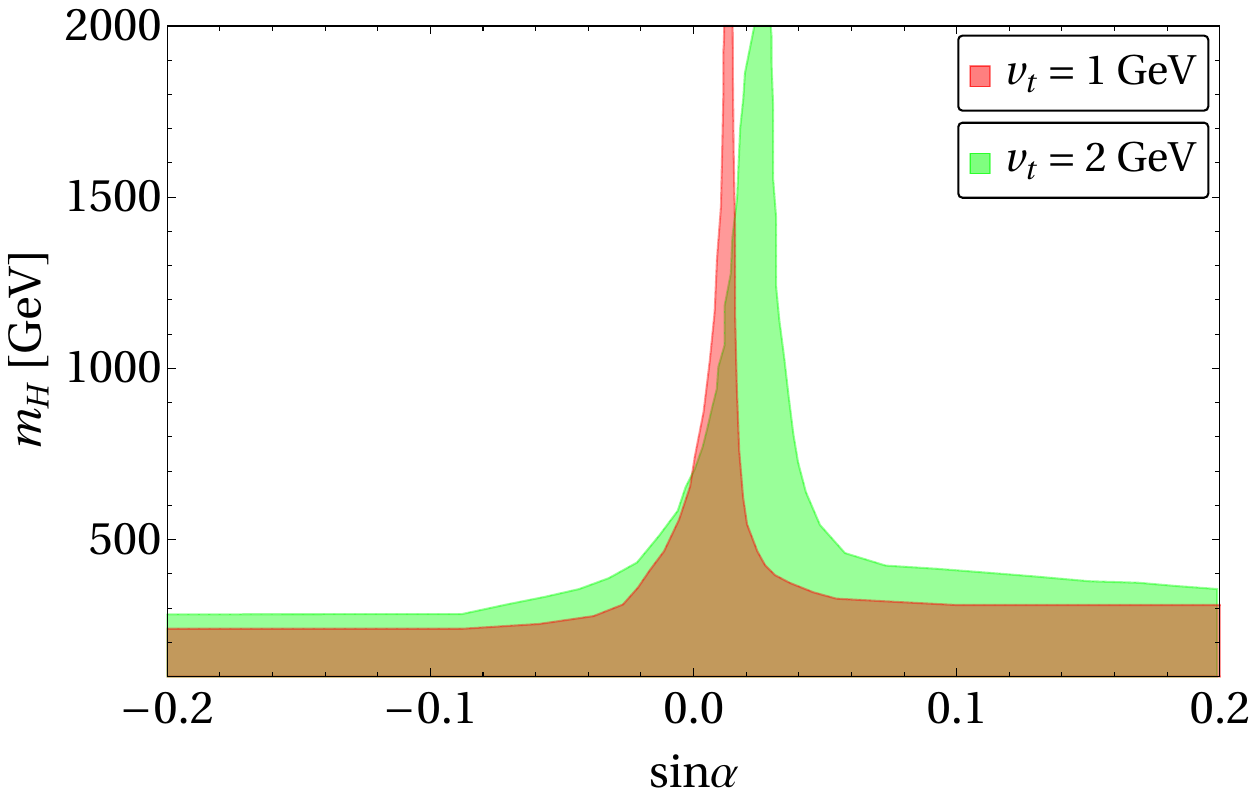}
\caption{Allowed regions from the combined constraints of unitarity and BFB in the limit $v_t\ll v$. The location of the narrow peak correspond
to the decoupling limit defined by \Eqn{e:sina}.}
\label{f:dec1}
\end{figure}

To illustrate the implications of the decoupling limit, we take the example of
$|y_2|\le 8\pi$ which, in terms of the physical parameters, reduces to
\begin{eqnarray}
\label{e:uniex1}
\left|\frac{1}{3}\left[m_5^2+ 2\left(m_h^2\sin^2\alpha+m_H^2\cos^2\alpha \right)
 \right] -m_3^2\cos^2\beta \right| \le 2\pi v^2\sin^2\beta \,.
\end{eqnarray}
In the decoupling limit when $v_t\ll v$, {\it i.e.}, $\sin^2\beta \ll 1$, the above
relation will be extremely constraining and will effectively reduce to the following
equality:
\begin{eqnarray}
\label{e:unieq1}
\frac{1}{3}\left[m_5^2+ 2\left(m_h^2\sin^2\alpha+m_H^2\cos^2\alpha \right)
\right] -m_3^2\cos^2\beta \approx 0 \,.
\end{eqnarray}
The conditions $|y_1|\le 8\pi$ and $|y_3-y_5|\le 16\pi$
will also have similar implications which we do not show
explicitly\footnote{The conditions
$|y_3|, \, |y_5|\le 8\pi$ can be combined to obtain $|y_3-y_5|\le 16\pi$ using the
triangle inequality.}.

\begin{figure}[!htb]
\centering
\includegraphics[width=0.48\textwidth,height=0.25\textheight]{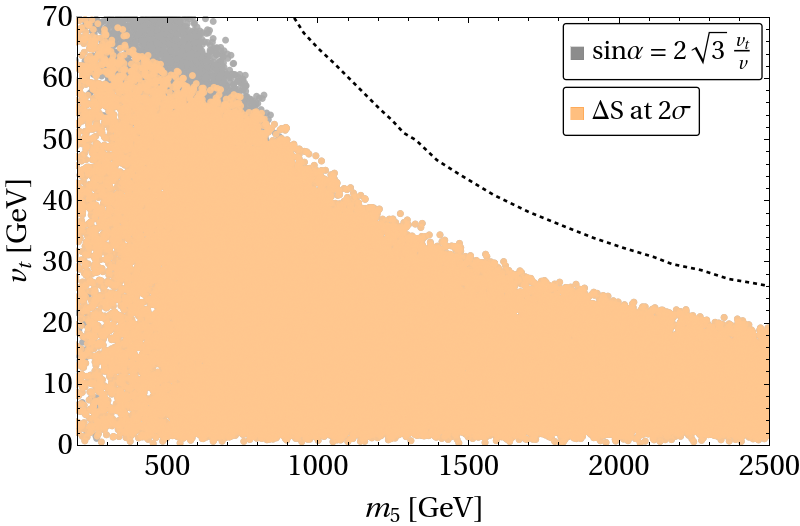}
\quad
\includegraphics[width=0.48\textwidth,height=0.25\textheight]{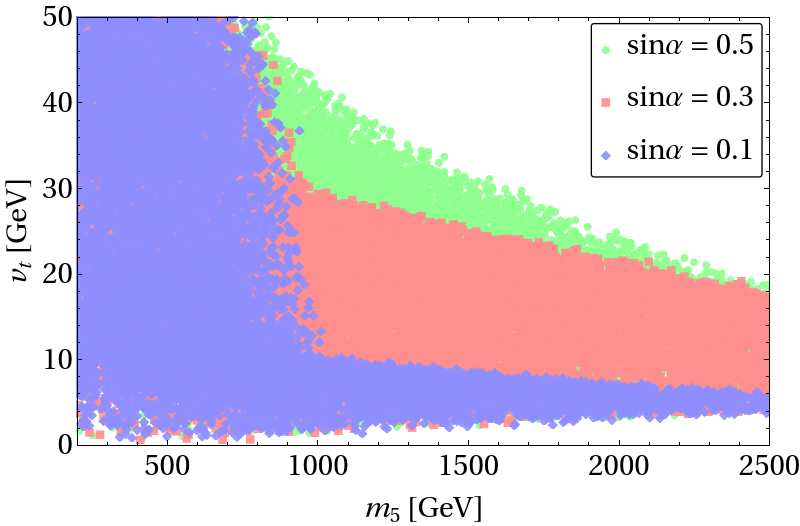}
\caption{Points allowed by the theoretical constraints of perturbative unitarity
and BFB constraints. In the left panel we have assumed 
\Eqn{e:sina1} to correlate $\sin\alpha$ and $v_t$. The dashed black line represents the
conservative bound given by \Eqn{e:LQT}. 
In the right panel different benchmark values of $\sin\alpha$ are chosen,
which are unrelated to $v_t$. The different colors in the right panel
correspond to different values of $\sin\alpha$ mentioned in the legends.
The gray points in the left panel are excluded when we impose the additional constraint
from $\Delta S$. As can be seen from the left panel, the constraint from $\Delta S$
starts to become relevant for $v_t \gtrsim 50$ GeV.  In the right panel, $\Delta S$ does
not impose any additional restriction for the displayed region of parameter space.
}
\label{f:reluni}
\end{figure}

Another type of constraint will arise from conditions like $|y_3|\le 8\pi$ which
reduces to
\begin{eqnarray}
\label{e:uniex2}
\left|m_3^2 -\frac{\sqrt{2}}{\sqrt{3}} \left(m_H^2-m_h^2\right)
\frac{\sin 2\alpha}{\sin 2\beta} \right| \le 4\pi v^2 \,.
\end{eqnarray}
Similar constraints can be obtained from $|y_4|, \, |y_5|\le 8\pi$. A common feature
of all these constraints is the occurrence of the ratio $(\sin 2\alpha/\sin 2\beta)$
which blows up in the limit $\sin\beta\ll 1$ and thus jeopardizes the unitarity
conditions for $m_H^2 \gg v^2$.  Therefore, imposition of the unitarity conditions
will entail a correlation between $\sin\alpha$ and $\sin\beta$ so that constraints
like \Eqn{e:uniex2} can be satisfied even for $v_t\ll v$. From the example
conditions of \Eqs{e:unieq1}{e:uniex2}, one may intuitively infer that the unitarity
conditions will be trivially satisfied for
\begin{subequations}
\label{e:dec}
\begin{eqnarray}
&&	\sin 2\alpha \approx \sqrt{\frac{3}{2}}\sin 2\beta \qquad {\rm with}~~ v_t\ll v
\label{e:sina} \\
{\rm and} && m_H^2 \approx m_3^2 \approx m_5^2 \approx \Lambda_1^2 \gg v^2 \,;
~~\Lambda_2^2 \ll v^2 \,. \label{e:M}
\end{eqnarray}
\end{subequations}
\Eqn{e:dec} defines the decoupling limit in the GM model. Using the definition of
$\tan\beta$ in \Eqn{e:tanb}, we may simplify \Eqn{e:sina} as
\begin{eqnarray}
\label{e:sina1}
\sin\alpha \approx 2\sqrt{3}\, \frac{v_t}{v} \,,
\end{eqnarray}
which will often be used as benchmark for our phenomenological analysis.
A visual confirmation of \Eqn{e:sina} is given in Fig.~\ref{f:dec1} where we see
that, for $v_t\ll v$, heavy nonstandard scalars beyond the TeV scale would require
$\sin\alpha$ to be strongly correlated to $v_t$. Such a correlation is not unique
to the GM model and can be found in the usual HTM as well\cite{Das:2016bir}. 

It should be noted that in the decoupling limit defined by \Eqn{e:dec}, the quartic
couplings of \Eqn{e:lambdas} take much simpler forms as follows:
\begin{eqnarray}
\label{e:dec-lam}
\lambda_1 \approx \frac{m_h^2}{8v^2} \,, \quad
\lambda_2\approx \lambda_5 \approx 0 \,.
\end{eqnarray}
Thus, only some of the quartic coefficients survive in the decoupling limit and $\lambda_1$ approaches the SM value.
Additionally, since the doublet-triplet mixing is also vanishingly small in
the decoupling limit, the physical scalar $h$, defined in \Eqn{e:h}, will have
SM-like couplings and can play the role of the SM-like Higgs boson observed at
the LHC\cite{ATLAS:2012yve,CMS:2012qbp}. Also comparing \Eqs{e:M1}{e:M} we see that a small triplet VEV
is intimately connected to large nonstandard scalar masses, which carries the
reminiscence of a type~II seesaw mechanism.

Another interesting point to note from \Eqs{e:Lam}{e:dec} is that the trilinear
coupling $M_1$ plays a crucial role in ensuring safe decoupling of the nonstandard
scalars. Therefore, variants of the GM model without the trilinear couplings $M_1$
and $M_2$, do not have a decoupling limit and therefore can be ruled out rather
easily\cite{Das:2018vkv}.

Next we display in Fig.~\ref{f:reluni} the points that pass the combined constraints
arising from unitarity and BFB. We exhibit results for the cases when \Eqn{e:sina1}
is satisfied as well as when $\sin\alpha$ and $v_t$ are unrelated. If $\sin\alpha$
and $v_t$ are unrelated then, as expected, the decoupling limit cannot be achieved
and as a result we see in the right panel of Fig.~\ref{f:reluni} that a substantial
area in the low $v_t$ region is excluded. We would like to comment here that our result agrees with
Ref.~\cite{Hartling:2014zca} with regards to the exclusion from the perturbative unitarity
and BFB requirements. The forbidden region in the upper right
corners of Fig.~\ref{f:reluni} can be qualitatively understood from the trilinear
couplings of the Higgs bosons with a pair of $W$-bosons. Let us express these
couplings as follows:\footnote{Note that $H_3^0$ being a pseudoscalar does not
possess coupling of the form $H_3^0W_\mu^+W^{\mu -}$.}
\begin{eqnarray}
\label{e:WWlag}
{\mathscr L}_{WWS}^{\rm tri} = gM_W\, W_\mu^+W^{\mu -}\left(\kappa_W^h h
 +\kappa_W^H H + \kappa_W^{H_5} H_5^0\right) + gM_W\, \frac{\kappa_2}{2}
 \left(W_\mu^+W^{\mu +}H_5^{--}+ {\rm h.c.} \right) \,,
\end{eqnarray}
where $S$ represents a generic scalar, $g$ denotes the $SU(2)_L$ gauge coupling strength
and the $\kappa$'s are given as,
\begin{subequations}
\label{e:kaps}
\begin{eqnarray}
	\kappa_W^h &=& \left(\sqrt{\frac{8}{3}}\sin\alpha\sin\beta +
	\cos\alpha\cos\beta \right)\,,   \label{e:kapWh} \\
        \kappa_W^H &=& \left(\sqrt{\frac{8}{3}}\cos\alpha\sin\beta -
	\sin\alpha\cos\beta \right)\,, \\
	\kappa_W^{H_5} &=& \frac{\sin\beta}{\sqrt{3}} \,, \\
	\kappa_2 &=& \sqrt{2}\sin\beta \,.
\end{eqnarray}
\end{subequations}
It is instructive to verify that these coupling modifiers obey the unitarity
sum rule\cite{Gunion:1990kf}
\begin{eqnarray}
\label{e:sum-rule}
\left(\kappa_W^h\right)^2 +\left(\kappa_W^H\right)^2 +\left(\kappa_W^{H_5}\right)^2
= 1 + \left(\kappa_2\right)^2 \,.
\end{eqnarray}
The Lee-Quigg-Thacker bound on the Higgs boson masses\cite{Lee:1977eg}, 
in this context, should thus read
\begin{eqnarray}
\label{e:LQT}
\left(\kappa_W^h\right)^2 m_h^2 +\left(\kappa_W^H\right)^2 m_H^2 +\left[
\left(\kappa_W^{H_5}\right)^2 +\frac{1}{2}\left(\kappa_2\right)^2 \right]m_5^2
\le 4\pi v^2 \,.
\end{eqnarray}
This inequality gives rise to a conservative bound from perturbative unitarity
which can qualitatively explain the forbidden region in the upper-right corner of Fig.~\ref{f:reluni}.
We show this bound as a black dashed line in the left panel of Fig.~\ref{f:reluni}.

To make the notion of decoupling more explicit, we calculate the trilinear couplings
of the SM-like Higgs, $h$, with a pair of charged scalars in the limit of \Eqn{e:sina}.
In particular, the factors that control the contributions of the charged scalar loops
in decays like $h\to\gamma\gamma$ and $h\to Z\gamma$ are given by\cite{Bhattacharyya:2014oka}
\begin{subequations}
\label{e:kapc}
\begin{eqnarray}
	\kappa_{3+} &\equiv& \frac{v}{2m_3^2} g_{hH_3^+H_3^-} \approx -\frac{1}{m_3^2}\left(m_3^2-\Lambda_1^2+m_h^2 \right) \,, \\
	\kappa_{5+} &\equiv& \frac{v}{2m_5^2} g_{hH_5^+H_5^-} \approx -\frac{1}{m_5^2}\left(2m_5^2-3m_3^2+\Lambda_1^2-\Lambda_2^2+m_h^2 \right) \,,  \\
	\kappa_{5++} &\equiv& \frac{v}{2m_5^2} g_{hH_5^{++}H_5^{--}} \approx  -\frac{1}{m_5^2}\left(2m_5^2-3m_3^2+\Lambda_1^2-\Lambda_2^2+m_h^2 \right)\,,
\end{eqnarray}
\end{subequations}
where \Eqn{e:sina} has been assumed. Clearly, when \Eqn{e:M} is also imposed, we will have
$\kappa_{3+}\,, \kappa_{5+}\,, \kappa_{5++}\approx 0$ implying that the charged scalars
are decoupled from the loop-induced Higgs decays in the limit of \Eqn{e:dec}, as expected.
In this context it should be emphasized that the expressions of \Eqn{e:kapc} crucially
depend on how the limit $\sin\alpha\to 0$ is approached. For example, instead of \Eqn{e:sina},
if we first apply $\sin\alpha\approx 0$ independent of $v_t$ and then take $v_t\ll v$, then
we would obtain
\begin{subequations}
	\label{e:kapc1}
	\begin{eqnarray}
	\kappa_{3+} &\approx& -1 \,, \\
	\kappa_{5+} &\approx&  -\frac{1}{m_5^2}\left(3m_3^2-2\Lambda_1^2 \right) \,, \\
	\kappa_{5++} &\approx& -\frac{1}{m_5^2}\left(3m_3^2-2\Lambda_1^2 \right)   \,,
	\end{eqnarray}
\end{subequations}
which do not lead to proper decoupling of the heavy charged scalars. Therefore, to ensure
safe decoupling of the nonstandard scalars one must approach $\sin\alpha\to 0$ limit in the
way dictated by \Eqn{e:sina}.

Here we wish to clarify the distinction between the notions of `alignment'
and `decoupling'. `Alignment' refers to the limit when the lighter neutral Higgs boson ($h$)
originates entirely from the $SU(2)_L$ doublet ({\it i.e.} $\sin\alpha=0$) making fermionic couplings of $h$ to be SM-like\footnote{
Putting $\sin\alpha =0$ in Eqs.~(\ref{lambda2}) and (\ref{lambda5}), we obtain $4\left(2 \lambda_2 - \lambda_5\right)v_t = M_1$. Note that, such a relation can also be inferred by demanding the off-diagonal element of the $2\times 2$ mass-matrix in the $h_d$-$H_5^{0\prime}$ basis ~\cite{Hartling:2014zca} to be zero. However, if $\sin\alpha =0$ limit is approached in this way, it will lead to alignment without decoupling.
}.
Additionally, in this limit $v_t\ll v_d$ so that the $SU(2)_L$ triplet Higgses have negligible
couplings of the form $S V_1^\mu V_{2 \mu}$, making the trilinear $hV_1^\mu V_{2\mu}$ couplings to be SM-like as well.
However, one of the crucial observations of our paper is that merely making
the tree level couplings of $h$ to be SM-like does not guarantee the decoupling of heavy scalars,
as has been emphasized through our Eqs.~(23) and (24).
For proper decoupling, we should approach $\sin\alpha\to 0$
and $\sin\beta\to 0$ in a correlated manner. Such discussions of alignment vs
decoupling are quite widespread for doublet extensions~\cite{Carena:2013ooa,Bhattacharyya:2014oka}.
In this work we have performed a similar analysis for the triplet extensions 
which has remained somewhat less explored in the literature so far.

We also wish to add that the discussion made in this section 
highlights an underemphasized fact that, in the scenario of superheavy nonstandard
scalars (much heavier than the EW scale) with only the SM-like scalar at the EW scale,
the perturbative unitarity constraints automatically push us towards
the decoupling limit (in a spirit similar to Ref.~\cite{Logan:2022uus}).
Such an aspect of perturbative unitarity has been discussed
earlier for nHDMs\cite{Bhattacharyya:2013rya,Bhattacharyya:2014oka,Bhattacharyya:2015nca} and the HTM\cite{Das:2016bir}.
Here we explicitly
demonstrate the connection between perturbative unitarity and the decoupling
limit for the GM model as well. We believe that this section will help to
view the perturbative unitarity constraints in a new light and serve as a motivation for
the parametrization that we advocate.

For later use, we also give the expression of the coupling modifier for the trilinear Higgs
self-coupling, which reads
\begin{eqnarray}
\label{e:klam}
\kappa_\lambda \equiv \frac{\lambda_{hhh}}{(\lambda_{hhh})^{\rm SM}} &=&
\cos^3\alpha\sec\beta +\frac{2\sqrt{2}}{\sqrt{3}}\sin^3\alpha\csc\beta +\frac{2\Lambda_1^2}{m_h^2}
\sin^2\alpha\cos\beta \left( \cos\alpha -\frac{\sqrt{2}}{\sqrt{3}} \sin\alpha\cot\beta \right) \nonumber \\ 
 && +\frac{\sqrt{2}}{3\sqrt{3}} \frac{\Lambda_2^2}{m_h^2}\sin^3\alpha\csc\beta \,.
\end{eqnarray}
One can easily check that $\kappa_\lambda=1$ in the limit of \Eqn{e:dec}.
Moving away from the decoupling limit, however, the deviation in $\lambda_{hhh}$ can
be significantly large depending on the parameter combinations.
Preliminary measurements of $\kappa_\lambda$ can already put
important constraints on the model parameter space, as will be discussed in more detail
in Sec.~\ref{s:direct}.

  In passing, we recall that the oblique $S$-parameter~\cite{Peskin:1991sw} has been known to put important
  constraints on the parameter space of the GM model~\cite{Gunion:1990dt,Englert:2013wga,Hartling:2014aga}.
  The new physics contribution to the $S$-parameter in the GM model is
  given by\cite{Hartling:2014aga}\footnote{It should be noted that this expression of $\Delta S$ relies
  on the assumption that the new physics scale is much larger than $M_Z$. Although it is possible
  to define the oblique parameters without this assumption\cite{Maksymyk:1993zm,Grimus:2008nb}, the corresponding
  expressions for the GM model do not seem to be available in the literature. Keeping this in mind,
  the results that follow from \Eqn{eq:spara} should be interpreted with caution.},
\begin{eqnarray}
  \Delta S &\equiv& S^{\rm GM}-S^{\rm SM} \nonumber \\
  &\approx& \frac{s_W^2 c_W^2}{\pi e^2}\left\{
  -\frac{e^2}{12 s_W^2 c_W^2} \left(\ln{m_3^2} + 5 \ln{m_5^2}\right)
  + 2|g_{ZhH_3^0}|^2 f_1(m_h,m_3)\right.\nonumber \\
  &&\left.+ 2|g_{ZHH_3^0}|^2  f_1(m_H,m_3) + 2\left(|g_{ZH_5^0H_3^0}|^2
          + 2|g_{ZH_5^+H_3^{-}}|^2\right)f_1(m_5,m_3)\right.\nonumber \\
  &&\left.+|g_{ZZh}|^2 \left[\frac{f_1(M_Z,m_h)}{2 M_Z^2}-f_3(M_Z,m_h)\right]
          -|g_{ZZh}^{\rm SM}|^2 \left[\frac{f_1(M_Z,m_h^{\rm SM})}{2 M_Z^2}
          - f_3(M_Z,m_h^{\rm SM})\right] \right.\nonumber \\
  &&\left.+|g_{ZZH}|^2 \left[\frac{f_1(M_Z,m_H)}{2 M_Z^2}-f_3(M_Z,m_H)\right]
          +|g_{ZZH_5^0}|^2 \left[\frac{f_1(M_Z,m_5)}{2 M_Z^2}-f_3(M_Z,m_5)\right] \right.\nonumber \\
  &&\left.+ 2|g_{ZW^+H_5^{-}}|^2 \left[\frac{f_1(M_W,m_5)}{2 M_W^2}-f_3(M_W,m_5)\right]\right\},
  \label{eq:spara}
\end{eqnarray}
where $e$ stands for the electric charge, $s_W (c_W)$ is the sine (cosine) of
the weak mixing angle, $M_Z$ is the $Z$-boson mass and $g_{XYZ}$ denotes the
coupling among the $X, Y$ and $Z$ particles excluding the Lorentz factor.
For explicit expressions of these couplings, we refer the reader to Ref.~\cite{Hartling:2014aga,Hartling:2014zca}
where a sign difference in the definition of $\sin\alpha$ needs to be taken into
account. The quantity $m_h^{\rm SM}$ is the reference value of the SM Higgs boson mass for which the
fit value of the $S$-parameter is obtained.
The $f_1$ and $f_3$ functions are given as,
\begin{align}
f_1(m_X,m_Y) = \left\{
                  \begin{aligned}
                    & \frac{1}{36(m_X^2-m_Y^2)^3} \left[ 5(m_Y^6-m_X^6) + 27 (m_X^4 m_Y^2-m_X^2 m_Y^4)
                       + 12 (m_X^6-3 m_X^4 m_Y^2) \ln{m_X} \right. \\
                    & \left. + 12(3 m_X^2 m_Y^4-m_Y^6)\ln{m_Y} \right], \quad \mathrm{for}~m_X \ne m_Y.\\
                    & \frac{1}{6} \ln m_X^2, \quad \mathrm{for}~m_X = m_Y.
                    \end{aligned}
                  \right.
\end{align}
and
\begin{align}
  f_3(m_X,m_Y) = \left\{
       \begin{aligned}
        & \frac{m_X^4 - m_Y^4 + 2 m_X^2 m_Y^2 \left(\ln m_Y^2 - \ln m_X^2\right)}
        {2 (m_X^2 - m_Y^2)^3}, \quad \mathrm{for}~m_X \ne m_Y.\\
        & \frac{1}{6 m_X^2}, \quad \mathrm{for}~m_X = m_Y.
        \end{aligned}
        \right.
\end{align}
At first glance the expression in \Eqn{eq:spara} might seem strange because of the
appearance of dimensionful quantities inside logarithms. However, the expression is meaningful
because $\Delta S$ remains unaffected if all the mass-dimensionful quantities are
multiplied by a common scale factor. This happens as a result of the following relation
satisfied by the couplings,
\begin{eqnarray}
  && -\frac{e^2}{s_W^2 c_W^2} + \frac{2}{3}\left\{|g_{ZhH_3^0}|^2 + |g_{ZHH_3^0}|^2 + |g_{ZH_5^0H_3^0}|^2
  + 2|g_{ZH_5^+H_3^-}|^2 \right\}
  + \frac{1}{3 M_W^2} |g_{ZW^+H_5^-}|^2 \nonumber \\
  && + \frac{1}{6 M_Z^2} \left\{|g_{ZZh}|^2 + |g_{ZZH}|^2 +|g_{ZZH_5^0}|^2 - |g_{ZZh}^{\rm SM}|^2 \right\} = 0
\end{eqnarray}  

For our numerical analysis, the fit value of $\Delta S$ has been taken to be\cite{Workman:2022ynf}
\begin{equation}
\Delta S = -0.01 \pm 0.07.
\end{equation}  
The impact of the $\Delta S$ constraints can be seen in the left panel of
Fig.~\ref{f:reluni} where we show the points excluded at 2$\sigma$ in gray. It can be
seen that the $\Delta S$ constraints start becoming important
for $v_t \gtrsim 50$ GeV. This observation holds
even when the benchmarks of $\sin\alpha$ are chosen independently of $v_t$.
Therefore, we do not explicitly show the excluded region in the right
panel of Fig.~\ref{f:reluni}. As we will see in Secs.~\ref{s:mu} and \ref{s:direct},
such large values of $v_t$ are already excluded by the collider constraints from the
LHC.

%
\section{Experimental constraints}
In this section we will describe the relevant experimental constraints
on the parameter space of the GM model. Firstly, we will discuss the
constraints coming from the measurement of the 125 GeV Higgs signal strengths.
After outlining the parameter space satisfying the theoretical constraints as described in
Sec.~\ref{s:uni} and the Higgs signal strength constraints, we will derive the
most relevant direct search bounds from the LHC on the remaining parameter region.
The future prospects for the HL-LHC will also be discussed\footnote{The indirect constraints
from flavor data (especially from $b\to s \gamma$) start becoming relevant
for $v_t \gtrsim 40$ GeV for all $m_3$ and $m_5$~\cite{Hartling:2014aga}.
As we shall see, this region of parameter space is excluded
by the direct LHC searches for nonstandard scalars.}.
\subsection{Constraint from Higgs signal strengths}
\label{s:mu}
The physical Higgs state $h$ with mass $m_h \approx$ 125 GeV in our scenario arises from the
mixing between two custodial singlets. This mixing will cause the couplings of
$h$ with the massive fermions and gauge bosons to deviate from
their corresponding SM values. Such deviations are tightly constrained by the
precision measurements of the Higgs boson couplings at the LHC.
Both the ATLAS~\cite{ATLAS:2021vrm} and CMS~\cite{CMS:2022dwd}
collaborations have studied various production and decay 
modes of the 125 GeV Higgs boson, thereby giving limits on the
signal strength observables defined as,
\begin{equation}
\mu^i_{j} = \frac{\sigma^i}{(\sigma^i)_{\rm SM}}\times \frac{{BR}_j}{({BR}_j)_{\rm SM}},
\end{equation}
where $\sigma^i$ represents the production cross section in the $i$-th mode and
${BR}_j$ denotes the branching ratio into the final state $j$.

\begin{figure}[!htb]
\centering
\includegraphics[width=0.48\textwidth,height=0.25\textheight]{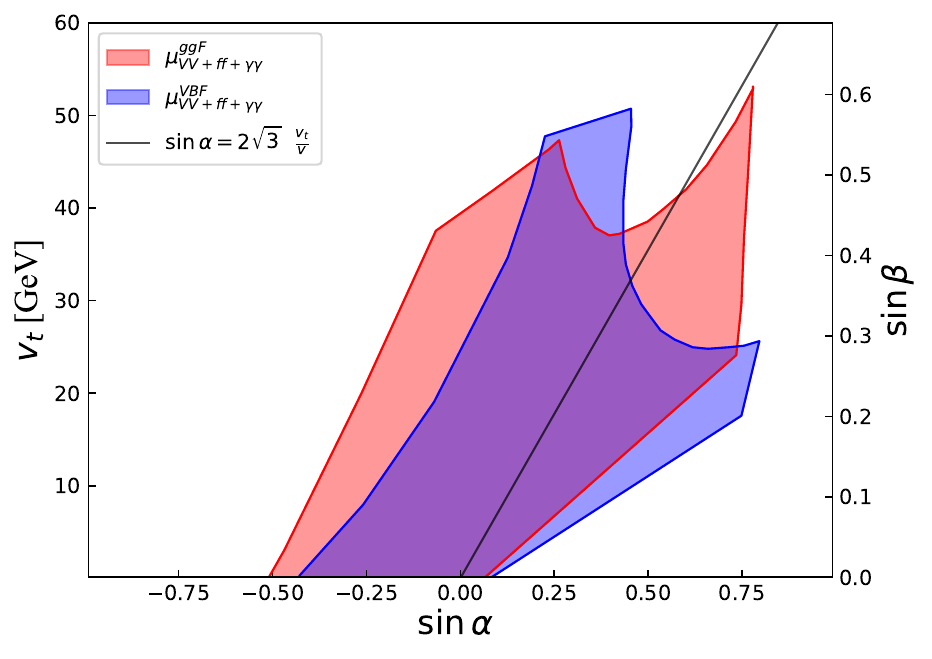}
  \caption{Allowed parameter space from Higgs signal strength measurements at 95\% confidence level (C.L.)~\cite{ATLAS:2021vrm,CMS:2022dwd}.
  For a fixed $\sin\alpha$, $v_t$ is restricted to a finite region, with clear upper and lower limits.}
 \label{higgsdata}
\end{figure}

We depict our results in Fig.~\ref{higgsdata}
which shows the range of allowed values for $v_t$ with respect to $\sin\alpha$, as obtained
after applying Higgs signal strength constraints.
We show the regions allowed by the signal strengths for
gluon-gluon fusion (ggF) and VBF production modes as red and blue shaded regions respectively.
We consider constraints from the $f\bar f$ and $VV$ final states,
$f$ and $V$ being the generic symbols for massive SM fermions and vector bosons respectively. 
Additionally, we also take into account the constraint from the $\gamma\gamma$ final state.
The allowed parameter space thus corresponds to the common region covered by the red and
blue shaded zones. The bottom line is that the Higgs signal strength data restricts $v_t$ to a finite region,
putting clearly-defined upper and lower limits on it for a fixed $\sin\alpha$.
%
%
\subsection{Direct search constraints from the LHC }
\label{s:direct}
As mentioned in the introduction, the presence of a rich variety of nonstandard scalars makes
the GM model subject to various constraints coming from the nonstandard Higgs boson searches at the
LHC. In this study we focus mostly on the potential impact of the bounds coming from the VBF production
of the custodial fiveplet charged and neutral scalars.
We also discuss direct search bounds on the custodial triplet and singlet nonstandard
Higgs bosons, namely $H_3^\pm$, $H_3^0$ and $H$. In the following, we briefly describe
the most important direct searches considered by us in this analysis.
\begin{itemize}
\item
The ATLAS collaboration has performed searches for 
the VBF production of a neutral heavy resonance $S_0$ decaying to $Z$-boson
pairs in the leptonic
final states~\cite{ATLAS:2020tlo}. The result is interpreted as a model-independent
upper bound on the production cross section times branching ratio $\left(\sigma_{\rm VBF} \times BR(S_0 \to ZZ)\right)$
for the VBF production process as a function of the
resonance mass. This data can be effective in constraining the neutral 
member of the custodial fiveplet $H_5^0$ as well as the custodial singlet $H$ in the GM model.

\item
  Searches for a charged Higgs boson have been performed in VBF production
  mode and its subsequent decay into $W^\pm Z$ modes~\cite{ATLAS:2022jho} and the null results have been
translated into exclusion bounds on the signal cross section as a function of the charged Higgs boson mass.
This bound can be crucial to constrain the properties of the charged Higgs state $H_5^\pm$.

\item
  The CMS collaboration has looked for the VBF production of a doubly charged scalar in
  like-sign $WW$ final states, producing model independent bounds on the corresponding signal strength
  as a function of doubly charged Higgs mass~\cite{CMS:2021wlt}. Additionally, the Drell-Yan production of a pair of
  doubly charged Higgs bosons with decays to $WW$ pairs has
  been investigated by the ATLAS collaboration~\cite{ATLAS:2021jol}.
  We employ these data to constrain the properties of the $H_5^{\pm\pm}$ particle.

\end{itemize}

Apart from the direct LHC searches, we also discuss here the constraints
coming from 125 GeV Higgs boson trilinear self-coupling measurements at the
LHC~\cite{ATLAS:2022jtk,CMS:2022dwd,Zabinski:2023jhr,ATLAS:2023qzf}.
Additionally, we checked our parameter space against the constraints
on the quartic gauge-Higgs coupling modifier $\kappa_{2V}$~\cite{Englert:2023uug}
from di-Higgs production process~\cite{ATLAS:2023qzf}.
This provides a much weaker bound compared to other constraints, excluding
only $\sin\alpha \gtrsim 0.81$. Therefore, we do not explicitly show this constraint
in the figures that follow.

We simulate the production of the nonstandard Higgs bosons
via the VBF process and their subsequent decays using \texttt{MadGraph-v-3.4.0}~\cite{Degrande:2011ua}.
For this, we implement our model in \texttt{FeynRules}~\cite{Christensen:2008py,Alloul:2013bka}
which generates \texttt{UFO} files to be used by {\tt MadGraph5\_aMC@NLO}.
The direct search limits are interpreted separately for the two mass hierarchies $m_5<m_3$ and $m_5>m_3$.

\begin{figure}[!htb]
		\centering
		\includegraphics[width=0.48\textwidth,height=0.25\textheight]{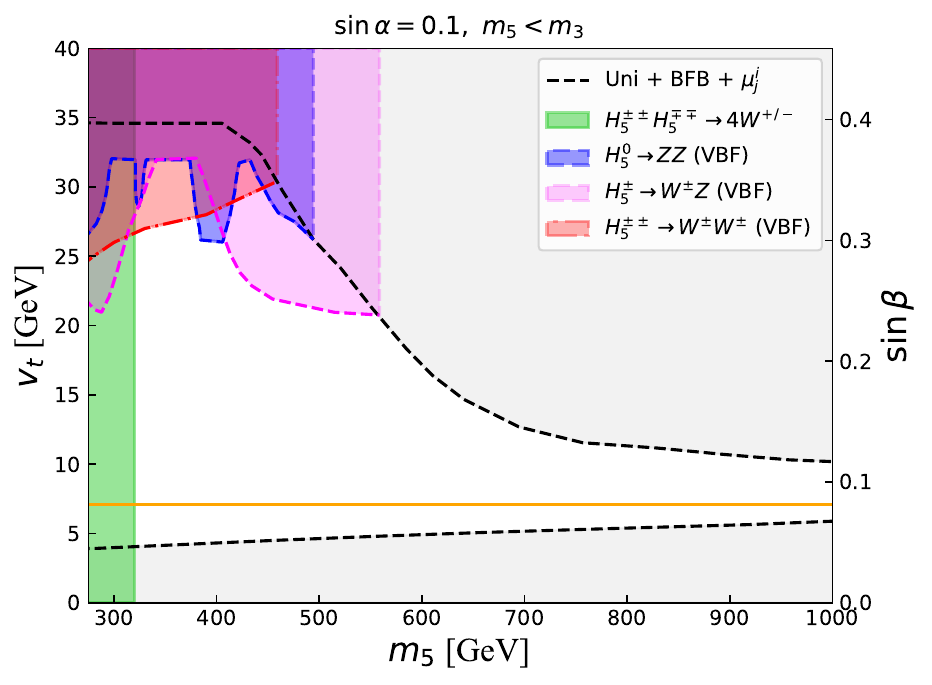}
		\includegraphics[width=0.48\textwidth,height=0.25\textheight]{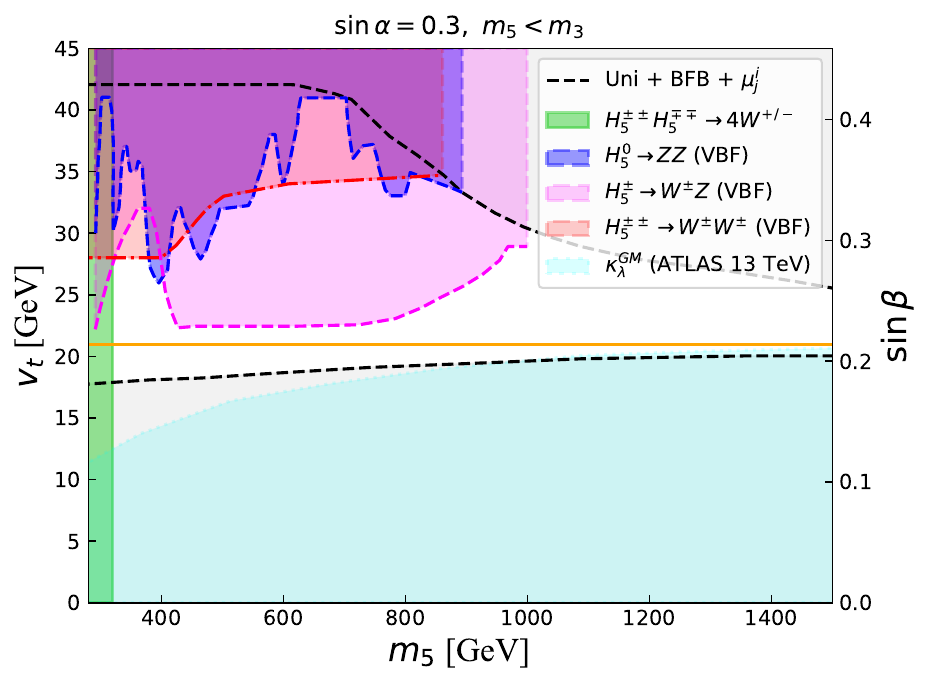}
		\caption{Combined theoretical and experimental constraints on the $m_5$-$v_t$ parameter plane for $m_5<m_3$ hierarchy
                  with $\sin\alpha = 0.1$ (left) and $\sin\alpha = 0.3$ (right).
                  The excluded regions are shaded with various colors explained in the text.
                  The orange solid line corresponds to the correlation $\sin \alpha = 2\sqrt{3}~v_t/v$.
                  Beyond the green band, the VBF searches dominate over Drell-Yan search channel
                  and put nontrivial constraints on the parameter space.}
		\label{comb-lim-m5ltm3}
	\end{figure}

\begin{figure}[!htb]
		\centering
		\includegraphics[width=0.48\textwidth,height=0.25\textheight]{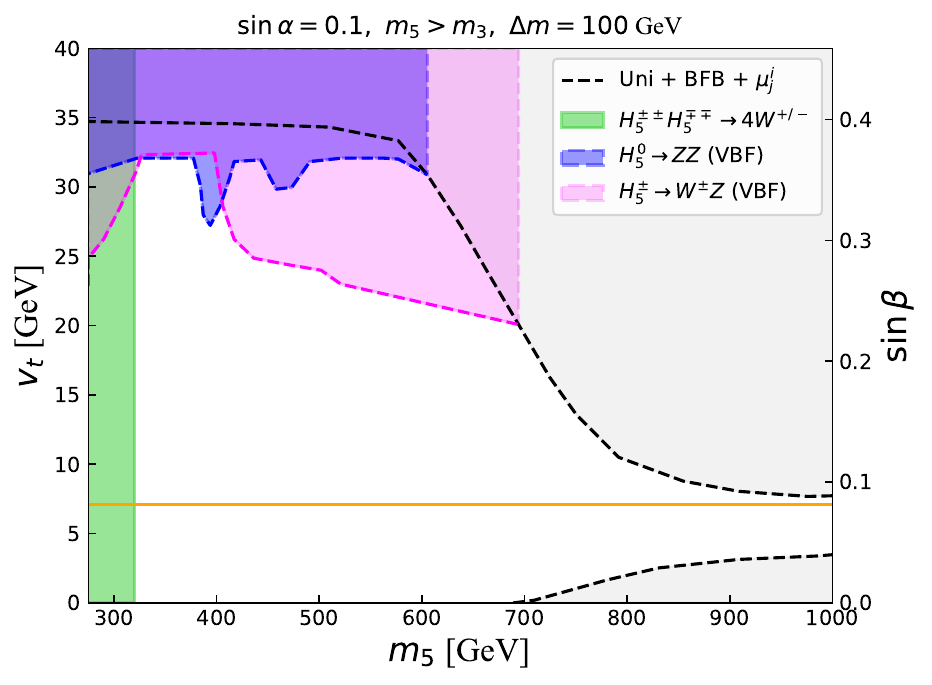}
		\includegraphics[width=0.48\textwidth,height=0.25\textheight]{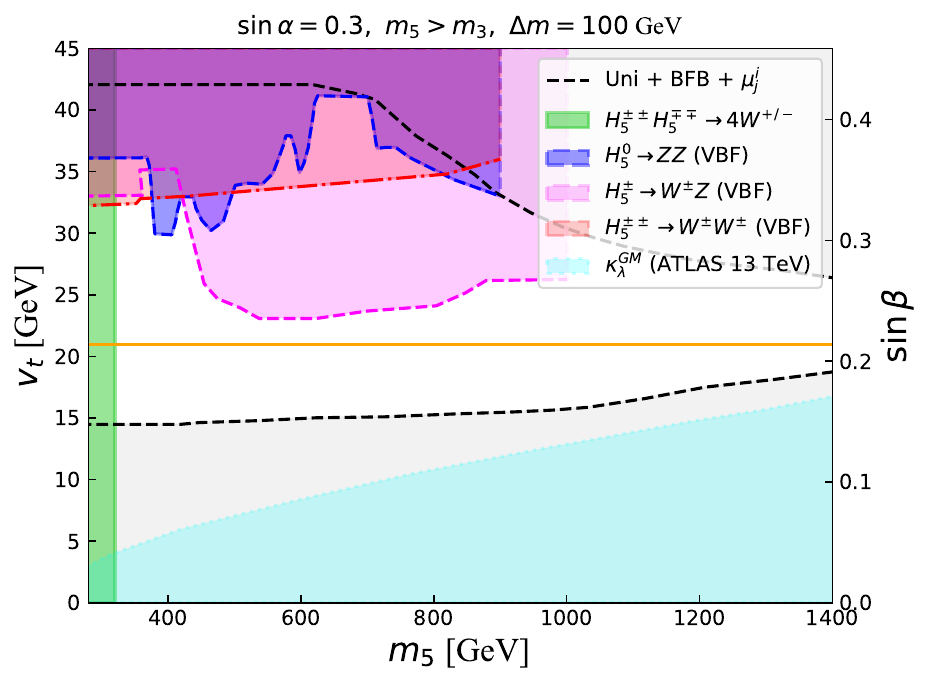}
		\caption{Combined theoretical and experimental constraints on $m_5$-$v_t$ parameter plane for $m_5>m_3$
                  hierarchy with $\Delta m = m_5 -m_3 = 100~\mathrm{GeV}$ for $\sin\alpha = 0.1$ (left) and $\sin\alpha = 0.3$ (right).
                  The excluded regions are shaded with various colors explained in the text. The orange solid line
                  corresponds to the correlation $\sin \alpha = 2 \sqrt{3}~v_t/v$.
                  The Drell-Yan search channels lose sensitivity beyond the green shaded region
                  where VBF constraints can still provide nontrivial constraints on
                  the parameter space.}
		\label{comb-lim-m5gtm3}
	\end{figure}

\begin{figure}
\begin{subfigure}[b]{0.52\linewidth}
    \centering{
 \includegraphics[height = 7.2 cm, width = 8.5 cm]{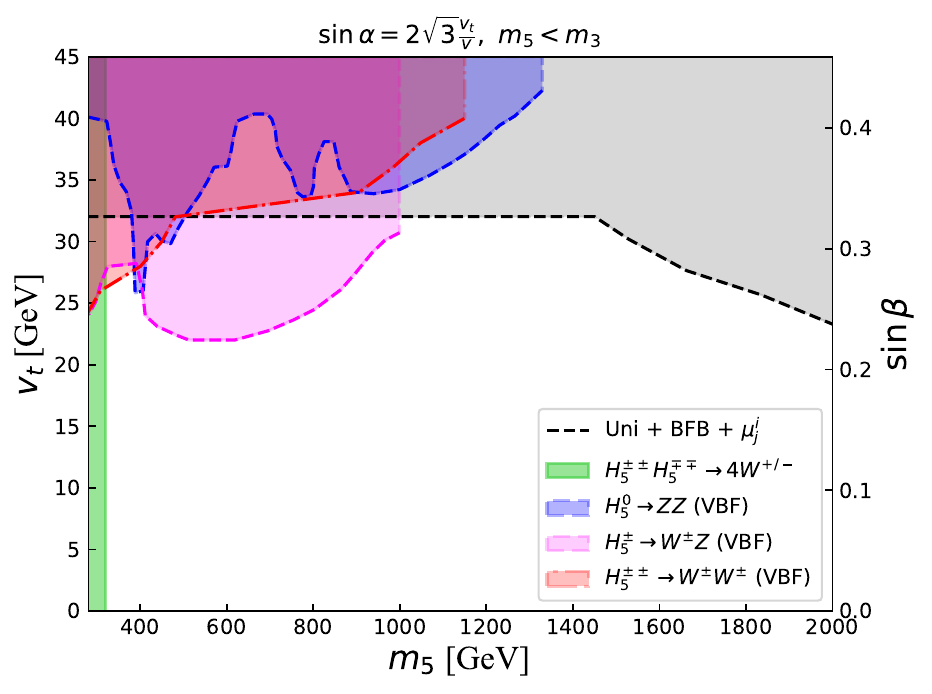}}
 \label{}
 \end{subfigure}
 ~
 \begin{subfigure}[b]{0.52\linewidth}
    \centering{
 \includegraphics[height = 7.2 cm, width = 8.5 cm]{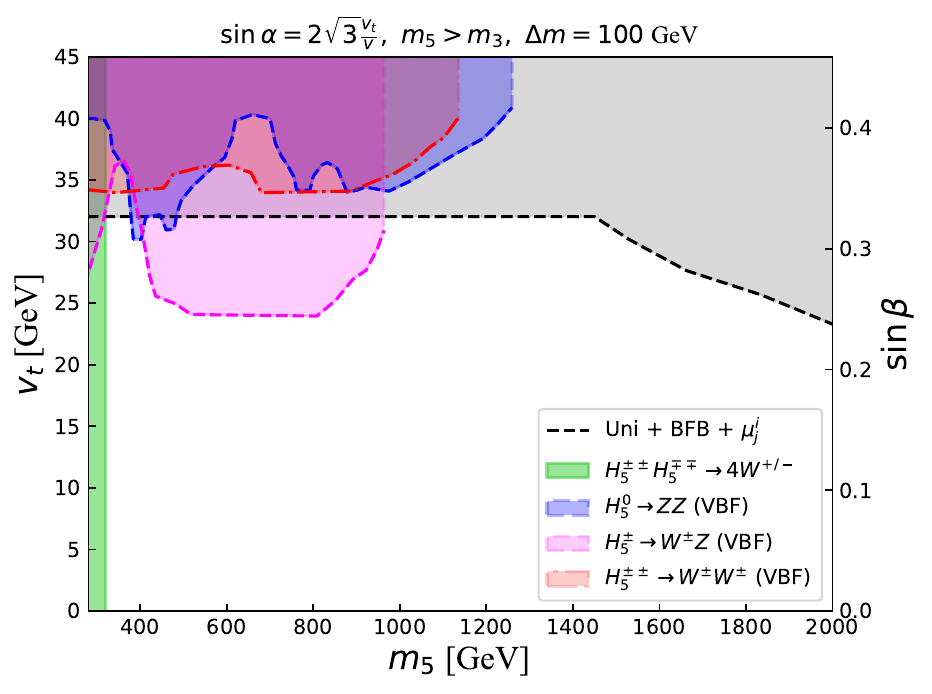}}
 \label{}
 \end{subfigure}
 \caption{Combined theoretical and LHC constraints on $m_5$-$v_t$ parameter plane for
    $\sin\alpha$ and $v_t$ correlated as in \Eqn{e:sina1}.
   The plots correspond to two hierarchies, $m_5<m_3$ (left) and $m_5 > m_3$
   with $(m_5 -m_3) = 100$~$\mathrm{GeV}$ (right).
   The excluded regions are shaded with colors detailed in the text.
   The Drell-Yan search channels lose sensitivity beyond the green band
   where VBF constraints can still provide effective constraints on
   the parameter space.
 }
\label{fig:collider_dependent}
 \end{figure}

In Fig.~\ref{comb-lim-m5ltm3} we show the combined effect of applying
the bounds from theoretical constraints, Higgs signal strength data and
the direct search limits from the LHC in the $m_5$-$v_t$ plane for the mass
hierarchy $m_5<m_3$. The gray
shaded region corresponds to parameter space excluded
from theoretical constraints of unitarity and BFB,
and the Higgs signal strength data. The collider bounds are then determined for the surviving parameter points.
The blue and purple shaded regions are excluded from direct search constraints on $H_5^0$
in the $ZZ$ decay mode~\cite{ATLAS:2020tlo} and $H_5^\pm$ in the $W^\pm Z$ final
states~\cite{ATLAS:2022jho} respectively.
The exclusion limit from the VBF production of $H_5^{\pm\pm}$ decaying to two like-sign
$W$ bosons~\cite{CMS:2021wlt} is shown as a red shaded region.
In Fig.~\ref{comb-lim-m5ltm3} , we depict our results for two benchmark values
of $\sin\alpha$. Larger values of $\sin\alpha$, independent of the value of $v_t$, correspond to farther
departure from the decoupling limit. Therefore, such $\sin\alpha$ values are expected
to receive more stringent constraints from the LHC searches in general.
As can be observed from the right panel of
Fig.~\ref{comb-lim-m5ltm3}, the allowed parameter space is squeezed
to a narrow region around the orange horizontal line, reinforcing the
strong correlation between $\sin\alpha$ and $v_t$ as dictated by \Eqn{e:sina1}\footnote{In Ref.~\cite{ATLAS:2022jho}
	the bound on the cross section has been reported for nonstandard masses
	up to 1~TeV. This is why the purple exclusion contours in Figs.~\ref{comb-lim-m5ltm3}, \ref{comb-lim-m5gtm3} and \ref{fig:collider_dependent}
	do not extend beyond 1~TeV.}.
We also show the exclusion limit from
the searches for Drell-Yan production of $H_{5}^{\pm\pm}$
decaying to $4W$ final state~\cite{ATLAS:2021jol}
as a green shaded region. However, the corresponding constraint
is much weaker than the others, with masses upto 320 GeV excluded independent of the value of $v_t$.
The limit from the associated Drell-Yan production of $H_5^{\pm\pm}H_5^{\mp}$~\cite{ATLAS:2021jol}
  is found to be even weaker. Beyond the green band, where the Drell-Yan search becomes ineffective,
  the VBF searches can still put important constraints on the parameter space. This is a common feature
  which can be seen in Figs.~\ref{comb-lim-m5gtm3} and \ref{fig:collider_dependent} as well.

Thus, one can observe from the Fig.~\ref{comb-lim-m5ltm3} that, while the theoretical constraints
can most effectively put bounds on the model parameter space towards the high $m_5$ region, complementary constraints
can be provided by the LHC VBF searches for the smaller values of $m_5$ as well.
With the advent of future colliders like the HL-LHC, the collider constraints are expected
to cut down deeper into the allowed parameter space towards lower $v_t$ region~\cite{Chiang:2021lsx}.
Thus, in the absence of any definitive signal of new scalars in the future,
triplet contribution to the electroweak VEV will become severely constrained.

In addition to direct collider bounds, we also take into account the
constraint coming from the measurement of 125 GeV Higgs self-coupling~\cite{ATLAS:2022jtk}.
We evaluate $\kappa_\lambda$ according to \Eqn{e:klam}.
The parameter space excluded from the $\kappa_\lambda$ constraint is shown as a cyan shaded region.
For $\sin\alpha = 0.1$ in the left panel of Fig.~\ref{comb-lim-m5ltm3},
the current measurement of 
$\kappa_\lambda$ does not impose any additional constraint on the parameter space.
On the other hand, only a marginal additional bound is obtained in the
high $m_5$ region for $\sin\alpha = 0.3$,
as can be seen from the cyan shaded region in the right panel of Fig.~\ref{comb-lim-m5ltm3}.
We would like to mention that $\kappa_\lambda$ constraint becomes increasingly more important for
larger values of $\sin\alpha~(> 0.3)$ and together with the bounds from unitarity and BFB, Higgs signal strength data and
direct LHC search limits, it can provide crucial constraints on the large $\sin\alpha$ 
parameter space.

We do not explicitly show the limits from $H_5^0 \rightarrow WW$ in the plots.
The bound presented for $H_5^0 \rightarrow WW$ via VBF production is much weaker than that
of $ZZ$~\cite{ATLAS:2022qlc}. Furthermore, we have checked that
this particular decay mode has half of the branching fraction to that of $H_5^0 \rightarrow ZZ$.
Thus, the bounds for $H_5^0 \rightarrow WW$ mode provides
no additional constraint on our parameter space.

In Fig.~\ref{comb-lim-m5gtm3} we show our allowed parameter region for
mass hierarchy $m_5 > m_3$ for two benchmark values of $\sin\alpha$. In this
case we present results for $\Delta m = m_5 - m_3 = 100$~GeV, which plays a crucial role in
determining the branching ratios of the decaying particles.  
 For $m_5<m_3$ scenario considered earlier,
the only possible decay modes for $H_5^\pm$ are those involving a pair of gauge bosons in the final state.
However, here the decay modes of $H_5^\pm$ become more diversified.
Additional decays into $H_5^\pm \to H_3^0 W^\pm$, $H_5^\pm \to H_3^\pm Z$ final
states are now kinematically accessible, which decreases the signal strengths for
$H_5^\pm\to W^\pm Z$ mode, weakening the corresponding bounds. Similarly, the presence
of $H_5^0 \to H_3^0 Z, H_3^\pm W^\mp$ modes results in the relaxation of the experimental constraint
on $H_5^0$. 
Our choice of $\Delta m = 100$~GeV serves as an illustrative benchmark to showcase the
relaxation of the collider constraints brought in by the opening up of the additional decay modes.

From Figs.~\ref{comb-lim-m5ltm3} and \ref{comb-lim-m5gtm3}, one may observe
a slight variation of the cyan region with $m_5$, as well as with the
hierarchy being considered. This may seem counterintuitive by looking at \Eqn{e:klam},
which shows no apparent functional dependence of $\kappa_\lambda$ on $m_5$. Such dependence of $\kappa_\lambda$ on $m_5$ is
an indirect effect generated by the perturbative unitarity and BFB conditions
which correlates the parameters $\Lambda_1$ and $\Lambda_2$ with values of $m_5$ and $m_3$.

In principle, the ATLAS data~\cite{ATLAS:2020tlo} should also translate
as a lower limit on $m_H$ as a function of $v_t$. Although the $H_5 Z Z$ and
$H Z Z$ couplings are of similar magnitudes, the width-over-mass
ratio for $H$ tends to become large (above 1\%) in a significant region of
parameter space, not respecting the narrow-width approximation.
On the other hand, the $BR(H\to ZZ)$
gets suppressed in this case because of the presence of various other
decays including di-Higgs and fermionic modes which were not present for $H_5^0$.
We have explicitly checked that, the effective cross section  $\sigma(pp \xrightarrow{\rm VBF} H \to ZZ)$
lies well below the ATLAS sensitivity reach. 
The limits on $H$ from $H\to hh$ searches can be effective
in constraining parameter spaces with $\sin\beta \gtrsim 0.4$~\cite{Ismail:2020zoz}\footnote{The custodial
symmetry forbids couplings of the form $H_5^0 hh$ and $H_3^0 hh$.
Also, $H_3^0$ being a pseudoscalar provides an additional reason not to give rise to the $H_3^0 hh$ coupling.}.
In our scans, this region lies inside the parameter space
already excluded from other complementary constraints.

For the $m_3<m_5$ hierarchy, the bounds on the charged Higgs boson $H_3^\pm$
must also be taken into account. Out of the two pairs of charged Higgs
bosons $H_3^\pm$ and $H_5^\pm$ of the GM model,
only the custodial triplet $H_3^\pm$ can couple to the SM fermions through
its mixing with the doublet. Thus, $H_3^\pm$ is likely to receive constraints
from the charged Higgs boson searches performed at the LHC.
The ATLAS collaboration has published search results for the production of charged Higgs bosons
decaying to $t\bar{b}$ final state~\cite{ATLAS:2018ntn}. 
The corresponding analysis by the CMS collaboration, however,
gives a much weaker bound~\cite{CMS:2019yat}.
In GM model the dominant production mode for $H_3^\pm$ is in association
with a $t\bar b$ pair. There are two possible decay modes for $H_3^\pm$
that dominates its total decay width. $BR(H_3^\pm \to t\bar b)$ is dominant
for smaller values of $m_3$. However, $BR(H_3^\pm \to W^\pm h)$ soon takes over, once it is
kinematically allowed~\cite{Ghosh:2019qie}. This relaxes the direct search bounds on $m_3$ from the LHC.
In fact, we have explicitly checked that our parameter space of interest lies
below the sensitivity region of ATLAS~\cite{ATLAS:2018ntn} in the $t\bar b$ final state.

The pseudoscalar $H_3^0$ may in principle be subjected
to the bounds coming from the LHC searches for
a CP-odd neutral scalar decaying into $Zh$ final state~\cite{ATLAS:2022enb}.
However, these searches target the production of the CP-odd state either
in ggF process or in association with b-quark pairs. Such production modes for $H_3^0$
suffer a $\mathcal{O}(\sin\beta)$ suppression in our scenario because of the doublet-triplet mixing,
making the bound considerably weak. We have checked that the signal yield for this process
stays below the limit for all $\sin\alpha$ and $v_t \lesssim 45$ GeV.

In Fig.~\ref{fig:collider_dependent}, we explain our results assuming the correlation
between $\sin\alpha$ and $v_t$
defined in \Eqn{e:sina1} to identify the decoupling limit of the model.
Compared to Figs.~\ref{comb-lim-m5ltm3}
and \ref{comb-lim-m5gtm3}, a significantly larger parameter space is now
allowed by the theoretical constraints of unitarity and BFB, especially in the low $v_t$ region.  
The current measurement of $\kappa_\lambda$ does not impose any additional constraints on the parameter space.
Once again, one can observe that stronger constraints from the direct collider searches
in future will drive $v_t$ to lower values, thereby constantly pushing us towards the decoupling limit.
\subsection{Future prospects: HL-LHC Projected Limits}
\label{s:future}
\begin{figure}[!htb]
	\centering
	\includegraphics[width=0.48\textwidth,height=0.25\textheight]{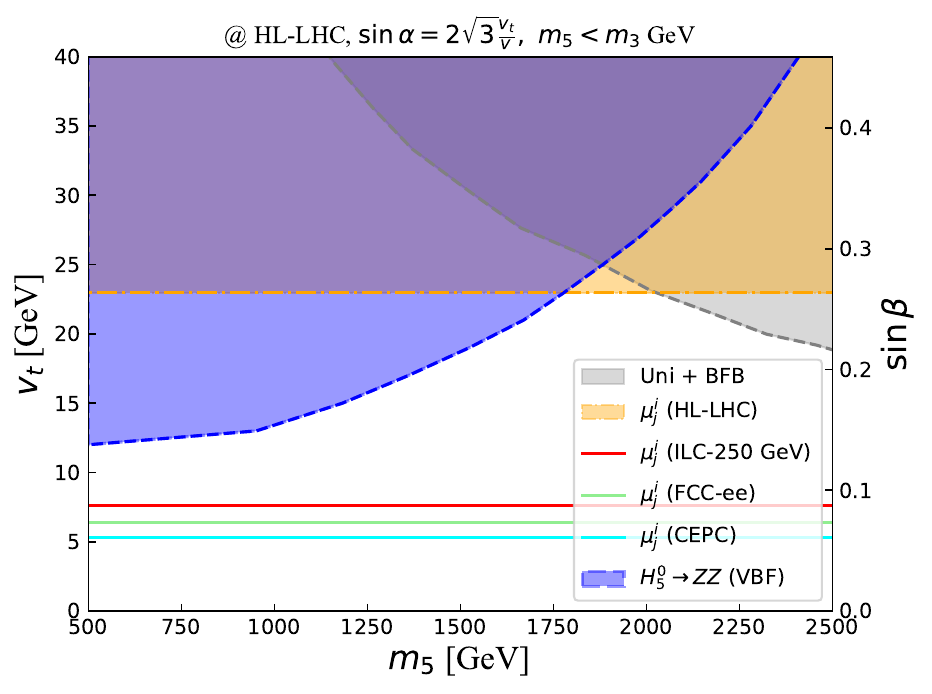}
	\caption{The expected sensitivity for the HL-LHC assuming the correlation $\sin \alpha = 2\sqrt{3}~v_t/v$.
          The blue shaded area represents the excluded regions from HL-LHC projected limit
          at $\sqrt{s} = 13~\mathrm{TeV}$ with integrated luminosity of 3000 $\mathrm{fb}^{-1}$ considering the
          VBF production of $H_5^0$ and its decay to $ZZ$~\cite{Cepeda:2019klc}.
          The orange shaded region corresponds to the expected exclusion reach from the HL-LHC
          measurements of the 125 $\mathrm{GeV}$ Higgs signal strength. The corresponding limits from
          the FCC-ee, CEPC and ILC are also shown. The region excluded from the
          unitarity and BFB constraints is shaded gray. The combined limits from HL-LHC and theoretical
        constraints will restrict $v_t$ to lower values, driving it closer to the decoupling limit.}
	\label{fig:future_collider_dependent}
\end{figure}

Here, we estimate the potential of the upcoming HL-LHC to probe the
parameter space of the GM model to a greater extent.
In Fig.~\ref{fig:future_collider_dependent} we present our results assuming the
correlation $\sin\alpha = 2\sqrt{3}~v_t/v$ for the $m_5<m_3$ mass hierarchy.
The projected sensitivity of HL-LHC searches for the VBF production of a BSM scalar resonance decaying
to  $ZZ$ mode~\cite{Cepeda:2019klc} is used to put constraints on the $m_5$-$v_t$ parameter plane.
The corresponding exclusion is shown as a blue shaded region. We also show the exclusion from
the unitarity and BFB constraints as a gray shaded region. The projected exclusion limit from the HL-LHC
measurements of Higgs signal strength is shown as the orange shaded region, assuming the central values of
the signal strengths to be consistent with the corresponding SM expectations.
Similar limits from
the planned Higgs factory experiments\cite{deBlas:2019rxi} like the Future Circular Collider (FCC-ee),
the Circular Electron-Positron Collider (CEPC) and the International Linear Collider (ILC) are also
shown in the plot.
For the direct searches,
one can see almost an order of magnitude improvement in the exclusion bounds compared to the current data,
pushing the parameter space down towards lower $v_t$ and hence towards the decoupling limit.
On the other hand, even with the projected
sensitivity of the HL-LHC to $\kappa_\lambda$, no additional constraints can be obtained on the 
allowed parameter space for correlated $\sin\alpha$ and $v_t$. The future Higgs factories can
be much more effective in this regard, restricting $v_t$ below 5 GeV.
The exclusion contours for the opposite mass hierarchy turn out to be
very similar to the one presented here.

\section{Summary}
\label{s:summary}
In this article, we studied the implications of the nonstandard scalar
(neutral as well as charged) searches in the VBF production channel at the LHC, taking the
GM model as an illustrative example. 
The trilinear coupling of a nonstandard scalar with a pair of massive SM gauge bosons
should come out to be proportional to the VEV of the electroweak multiplets from which
the nonstandard scalar is primarily derived.
However, as outlined in the introduction, for models comprising of only $SU(2)_L$
doublets in the scalar sector, the Higgs signal strength measurements force
us towards the alignment limit when the SM-like Higgs scalar arises from
a particular $SU(2)_L$ doublet in the Higgs basis~\cite{Branco:2011iw,Bhattacharyya:2015nca},
which consumes the entire electroweak VEV.
Thus, for these multi Higgs-doublet models, the trilinear couplings mentioned above
become extremely suppressed near the alignment limit.
Consequently, the VBF searches for the nonstandard scalars will not be very effective in constraining
these multi Higgs-doublet scenarios. 

For models featuring higher $SU(2)_L$ multiplet scalars (in addition to the SM Higgs-doublet),
the tree level value of the electroweak $\rho$-parameter can get significantly modified.
Therefore, the VEV of the additional multiplets may get severely constrained from the precision electroweak measurements.
The HTM
is a prime example of this scenario when the triplet VEV can, at the most, be of ${\cal O}(1\rm~GeV) $.
As a result, the trilinear couplings
will still be too small to be impactful in the VBF searches for the nonstandard scalars.
One may accommodate larger VEVs for the higher $SU(2)_L$
multiplets either by restoring the custodial symmetry~\cite{Kundu:2021pcg} or by arranging
accidental cancellations~\cite{Chiang:2018irv}. The GM model constitutes an example of the
first category where the VEV of the $SU(2)_L$ triplets can be
substantially large while still maintaining $\rho = 1$ at the tree-level. 
Thus, it serves as a well-motivated example where the VBF searches can put
important constraints on the parameter space especially on the triplet VEV as a function of the common 
mass of the members of the custodial fiveplet.

In our study of the GM model, we have first analyzed the theoretical constraints from unitarity and BFB.
Here we observe that, for low values of the triplet VEV, we need to have a correlation
between $\sin \alpha$ and $v_t$ to allow for very heavy nonstandard scalars decoupled from the electroweak scales.
The correlation $\sin \alpha \simeq 2\sqrt{3}~v_t/v$ grows stronger as the
ratio $v_t/v$ becomes smaller. Such a feature seems to have been
under-emphasized in the literature.
Because of this correlation, we systematically approach the SM limit first by imposing 
$\sin \alpha = 2\sqrt{3}~v_t/v$ and then requiring $v_t \ll v$.
 We also take into account the constraints coming from the
 125 GeV Higgs signal strength measurements.
 
Next, we study the phenomenological constraints arising from the VBF searches for
the nonstandard scalars.  Since
the members of the custodial fiveplet are derived
solely from the component fields of the $SU(2)_L$ triplets (see \Eqn{eq:ch}),
they do not couple to the SM fermions at all and therefore can be exclusively
probed via the VBF channels. 
We have primarily considered the direct searches in the $pp \xrightarrow{\rm VBF} S \to V_1V_2$ channel.
We have explored benchmark scenarios with uncorrelated $\sin \alpha$ and $v_t$ 
as well as with $\sin \alpha = 2\sqrt{3}~v_t/v$.
We have found that the LHC constraints can be complementary to the theoretical constraints from unitarity
and BFB to constrain the triplet VEV as a function of the custodial fiveplet mass.
For $m_5 < m_3$, the upper bound on the triplet VEV can be as strong
as $v_t \lesssim 25~\rm GeV$.
The relaxation of this bound has also been demonstrated for the opposite hierarchy, $m_5 > m_3$.
With the improved sensitivity projected for the HL-LHC, 
more stringent constraints on the triplet VEV are expected from the VBF searches,
implying that the combination of the direct HL-LHC collider limits and the theoretical constraints will 
enforce the decoupling limit of the GM model if no deviation from the SM is found in future.
Our analysis thus goes to show that the VBF searches for nonstandard scalars can be really useful
in restricting the non-doublet contributions to the electroweak VEV and thereby providing valuable
intuitions into the constructional aspects of new BSM scenarios. This observation, in turn,
underscores the fact that the null results for the BSM searches at the LHC are a lot more than just
upper bounds on cross sections as they can be translated into practical insights regarding the
anatomy of EWSB. 

\section*{Acknowledgements}
DD thanks the Science and Engineering Research Board, India for financial support
through grant no. CRG/2022/000565.
IS acknowledges the support from project number RF/23-24/1964/PH/NFIG/009073 and from DST-INSPIRE, India, under grant no. IFA21-PH272.
NG would like to thank Harish-Chandra Research institute for hospitality and computation facility while the project was going on. NG would also like to acknowledge IOE-IISC fellowship program for financial support. SM acknowledges Infosys-TIFR leading edge research grant. We also thank the International Centre for
Theoretical Sciences (ICTS) for their warm hospitality during the discussion meeting - ``Particle Physics: Phenomena, Puzzles, Promises'' (code: ICTS/p2p32022/11) where a significant progress of this work has been made.

\bibliographystyle{JHEP}
\bibliography{ref.bib}


\end{document}